# Simultaneous transport and tunneling spectroscopy of moiré graphene: Distinct observation of the superconducting gap and signatures of nodal superconductivity

Jeong Min Park[1,2,†,*], Shuwen Sun[1,†], Kenji Watanabe[3], Takashi Taniguchi[3], and Pablo Jarillo-Herrero[1,*]

[1]Department of Physics, Massachusetts Institute of Technology; Cambridge, Massachusetts 02139, USA.
[2]Department of Physics, Princeton University; Princeton, New Jersey 08544, USA.
[3]National Institute for Materials Science; Namiki 1-1, Tsukuba, Ibaraki 305-0044, Japan.

*Corresponding author. Email: parkjane@princeton.edu (J.M.P); pjarillo@mit.edu (P.J.H)
†These authors contributed equally to this work.

**Understanding the nature of superconductivity in magic-angle graphene remains challenging. A key difficulty lies in discerning the different energy scales in this strongly interacting system, particularly the superconducting gap. Here, we report the first simultaneous tunneling spectroscopy and transport measurements of magic-angle graphene, providing a novel approach to probe the superconducting state. This approach allows us to identify two coexisting V-shaped tunneling gaps with different energy scales: a distinct low-energy superconducting gap that vanishes at the superconducting critical temperature and magnetic field, and a higher-energy pseudogap. The superconducting tunneling spectra display a linear gap-filling behavior with temperature and magnetic field and exhibit the Volovik effect, consistent with a nodal order parameter. Our work reveals the unconventional nature of the superconducting gap in magic-angle graphene and establishes an experimental framework for multidimensional investigation of tunable quantum materials.**

Understanding the pairing symmetry of superconductors is of fundamental interest in condensed matter physics. A key aspect is the accurate experimental identification of the superconducting gap through techniques sensitive to the system's density of states (DOS). However, distinguishing the superconducting gap from other types of energy gaps poses significant challenges. A notable example of this is the coexistence[1,2] of the superconducting gap and the pseudogap[3–5] in cuprate superconductors, where the pseudogap may not necessarily reflect the nature of the superconducting gap. To address these challenges, a multifaceted measurement approach is desirable.

This approach is particularly relevant in the study of recently discovered moiré superconductors[6], where the close proximity between multiple correlated states in their phase diagram, and their sensitivity to disorder, lead to difficulty in correlating results obtained by different techniques on different devices. For example, experiments on the magic-angle graphene family[7–11], characterized by alternating twist angles[12] of $\theta$ and $-\theta$, suggest an unconventional nature of its superconducting phase[7–11,13–22]. However, a direct spectroscopic identification of the superconducting gap and its correlation with transport properties has not been performed.

In this article, we introduce a novel experimental scheme based entirely on van der Waals (vdW) materials, which enables the simultaneous investigation of tunneling spectroscopy and transport phenomena[23], including variable temperature and magnetic field dependence measurements. We use this technique to establish a direct correlation between the tunneling density of states and the dissipationless transport properties of magic-angle twisted trilayer graphene (MATTG)[8,9]. Notably, the tunneling spectra reveal the coexistence of two distinct V-shaped gaps with different energy scales. Both gaps exhibit a similar doping dependence as the superconducting critical temperature; however, their responses to temperature and magnetic field are markedly different. This distinction allows us to unambiguously identify the low-energy tunneling gap as the superconducting gap. Furthermore, we find that the superconducting gap exhibits a sensitive response to temperature and magnetic field, both in parallel and perpendicular directions to the two-dimensional plane, suggesting a nodal superconducting order parameter.

Our experimental setup with the vdW heterostructure is illustrated in Fig. 1a. It consists of the following layers from top to bottom: a top graphite gate, a hexagonal boron nitride (hBN) dielectric, a top MATTG (t-MATTG), a three-atomic-layer hBN tunnel barrier, a few-nanometer-thick hBN spacing layer with a hole, a bottom MATTG (b-MATTG), another hBN dielectric, and a bottom graphite gate. Both the t-MATTG and b-MATTG layers are shaped into Hall bars, each independently contacted with five probes to enable comprehensive transport measurements. The top and bottom gates control the electrostatic doping of the MATTG layers. When a dc bias voltage $V_{int}$ is applied across the MATTG layers, electrons can tunnel between them, with the tunneling current being proportional to the DOS in the system. Using this setup, we can perform four-probe tunneling spectroscopy. The strategically placed hole (see inset in Fig. 1a) confines the electron tunneling to the core region of the four-probe transport measurements. This design, with a fixed tunneling region, enables precise comparisons between transport and tunneling features, as a function of continuous temperature and magnetic field for the first time in moiré materials.

Furthermore, employing a MATTG layer to probe another MATTG layer ensures a matched work function, providing accurate tunneling spectra.

**Simultaneous transport and tunneling spectroscopy**

We first perform four-probe transport measurements separately on each of the two MATTG layers while varying the top gate ($V_{TG}$) and bottom gate ($V_{BG}$) voltages. Figure 1b shows the longitudinal resistance of t-MATTG, $R_{xx,Top}$. The features are primarily dependent on $V_{TG}$, as the metallic states in b-MATTG screen the effect of $V_{BG}$. The resistive states around $V_{TG} \approx \pm 7.5$ V (when $V_{BG} = 0$ V) indicate full filling of the flat bands, typically denoted as $\nu = 4n/n_s = \pm 4$, where $n$ is the carrier density and $n_s = 8\theta^2/(\sqrt{3}a^2)$ is the superlattice density ($a = 0.246$ nm is the graphene lattice constant). We observe superconductivity in the doping ranges $V_{TG} = -6.1$ V~$-3.5$ V and 3.4 V~6 V on the hole-doped and electron-doped sides, respectively, corresponding to the $\nu = -2 - \delta$ and $\nu = +2 + \delta$ domes, where $\delta$ is the additional electrostatic doping beyond $\nu = \pm 2$. For b-MATTG (Fig. 1c), the features are mostly dependent on $V_{BG}$, with the superconducting domes emerging from $V_{BG} \approx \pm 6.5$ V when $V_{TG} = 0$ V. For clarity, the raw gate voltages are consistently presented throughout this article.

Having confirmed that both MATTG layers exhibit tunable phases including superconductivity, we perform tunneling spectroscopy, with b-MATTG gated to a normal metallic phase ($V_{BG} = -3$ V). This configuration allows the measurement of the DOS continuously across the t-MATTG phase space, using b-MATTG as a tunneling probe. Figure 1d shows the tunneling conductance $dI/dV$ as a function of $V_{int}$ and $V_{TG}$. The tunneling spectra show Van Hove singularities (VHS) which are sharp outside the flat bands and become broader when the chemical potential is within the flat bands, also exhibiting the cascade of phase transitions near integer filling $\nu$ of MATTG, consistent with previous measurements[24–30].

As we zoom into smaller energy scales, further intricate phenomena are revealed. Figure 1e shows this parameter space within an energy scale of less than 0.5 meV. One of the most prominent features is that $dI/dV$ develops a dip in two specific ranges of doping, indicating a depletion in the DOS, which signals the presence of a gap. Remarkably, this range corresponds exactly to where the transport resistance becomes zero, i.e. a superconducting phase, as shown in Fig. 1f. This marks the first direct correlation between a tunneling gap and a superconducting state measured by transport in a moiré superconductor. Examining the $dI/dV$ spectra as a function of doping (Fig. 1g, h; with the corresponding gate ranges marked by matching color bars in Fig. 1f), V-shaped tunneling spectra are revealed in the superconducting phase. The contrasting flat DOS beyond the full filling confirms that the b-MATTG metallic probe does not contribute any additional features. A complementary measurement, using the metallic phase in t-MATTG outside the flat band ($V_{TG} = -10$ V) to probe b-MATTG (see Fig. S1), shows the V-shaped curves that appear exclusively in the superconducting phase of b-MATTG, yielding similar results.

**Coexistence of two distinct gaps**

Figure 2a shows a plot of $dI/dV$ versus $V_{TG}$ for an intermediate $V_{int}$ range. In the doping ranges where the superconducting phase exists (Fig. 1f), we observe the low-energy-scale depletion of DOS corresponding to the V-shaped gaps identified in Fig. 1g and 1h, shown as the dark blue features around zero bias in Fig. 2a. However, in addition, another broader and shallower diamond-shaped gap feature is also present, albeit with a different energy scale of a few meV. To gain further insights, we perform tunneling spectroscopy as a function of temperature over this energy scale, as shown in Fig. 2b. At the lowest temperature (0.11 K, black curve), the tunneling spectrum shows a remarkable coexistence of two gaps: a sharp low-energy gap and a broader higher-energy gap, $\Delta_{HG}$. As we increase the temperature, the tunneling DOS at low energy rapidly fills up on the scale of the superconducting critical temperature $T_c \sim 1$ K (see Fig. 4c for transport measurements and Methods for definition), while the spectrum associated with $\Delta_{HG}$ barely changes. If we further increase the temperature, the $\Delta_{HG}$ spectrum changes at a slower rate, persisting up to a much higher temperature $T_{HG} \sim 8$ K (Fig. 2c), where $T_{HG}$ is the approximate temperature at which $\Delta_{HG}$ vanishes. The different behaviors of these two gaps can be seen in Fig. 2c inset, which shows the temperature dependence of the minimum tunneling conductance $(dI/dV)_{min}$. A clear change in the slope is observed around $T_c$. Figures 2d-e show the perpendicular magnetic field $(B_\perp)$ dependence of the $dI/dV$ spectrum. The low-energy gap rapidly fills up on the scale of the critical magnetic field $(B_{c,\perp})$ measured in transport (see Fig. 5 and Fig. S6). Interestingly, $\Delta_{HG}$ is barely affected by $B_\perp$, even at fields much greater than $B_{c,\perp}$. These behaviors are consistent on the hole-doped side as well, as demonstrated in Fig. S2. From our observations, we conclude that the 'inner', low-energy, tunneling gap corresponds to the superconducting gap, since (1) it is present only in the doping ranges where superconductivity appears in transport; (2) it vanishes at the critical temperature determined by transport; and (3) it vanishes at the critical perpendicular magnetic field determined by transport. On the other hand, the higher-energy gap has a different nature: although it is also present only in the doping ranges where superconductivity is seen in transport, it survives to much higher temperature and magnetic field, and hence $\Delta_{HG}$ is not the superconducting gap. This higher-energy gap may be a pseudogap (e.g. a large pairing gap amplitude due to pre-formed pairs) or another gap whose origin needs to be determined. Our experiments are the first to report the coexistence of two distinct energy gaps in the tunneling spectroscopy of magic-angle graphene: the superconducting gap and a higher-energy gap. These findings may help clarify a puzzling observation in previous scanning tunneling microscope (STM) studies on magic-angle graphene, where tunneling spectra in the doping range corresponding to the superconducting phase did not exhibit the expected temperature and magnetic field dependence[13,14].

**Nodal nature of superconducting gap**

Having established the superconducting tunneling gap distinct from $\Delta_{HG}$, we undertake a more detailed investigation of the superconducting phase near optimal doping on the hole-doped regime ($V_{TG} = -4.55$ V). Figure 3a shows the temperature dependence of the superconducting gap tunneling spectra from 120 mK to 2.2 K, while Fig. 3c shows that of transport resistance at the same doping. In the latter, the onset of the superconducting transition begins around $T = 2$ K with a BKT transition temperature of $T_{BKT} \approx 0.86$ K. In order to analyze the shape of the superconducting tunneling spectrum, we normalize it by a spectrum at higher temperature to minimize the $\Delta_{HG}$ contribution[31]. We choose $T = 1.4$ K (slightly above $T_c$) as the normalization temperature. Figure 3b shows the normalized tunneling conductance at $T = 120$ mK. We fit the spectrum with a modified Dynes formula[32,33], incorporating a finite gap distribution to account for a potential twist angle inhomogeneity[34] (see Methods). We find that the fitting works the best when a nodal order parameter is used (see Fig. S4 for a comparative analysis with the *s*-wave fitting). The mean value of the superconducting gap is $\bar{\Delta}_{SC} = 0.159$ meV (depending on the normalization temperature $T = 1.2$ K$\sim$1.6 K for the tunneling spectra, the mean value of the gap may vary in an approximate range of $\bar{\Delta}_{SC} = 0.149$ meV$\sim$0.164 meV), and by direct comparison with transport measurements, we obtain the ratio $2\bar{\Delta}_{SC}/k_B T_{BKT} \approx 4.3$. We note that this gap size is smaller than, but close to, the gap scales obtained by Andreev reflection measurements on MATBG[13] and MATTG[14].

To further investigate the behavior of the superconducting gap, we fit the normalized tunneling spectra at different temperatures (Fig. 3d, dashed lines) and extract the temperature dependence of $\bar{\Delta}_{SC}$. The raw values of the zero bias conductance $dI/dV$ ($V_{int} = 0$) and the extracted $\bar{\Delta}_{SC}$ versus $T$ are summarized in Fig. 3e. Remarkably, $dI/dV$ ($V_{int} = 0$) immediately rises with increasing $T$, well before $T$ reaches $T_{BKT}$. There is a transition near $T_c$, where the increasing rate becomes smaller as the superconducting gap disappears and $\Delta_{HG}$ remains, as previously seen in Fig. 2c inset. On the other hand, $\bar{\Delta}_{SC}$ initially decreases slowly with $T$, while it decreases faster as $T$ approaches $T_c$. Such 'gap filling' behavior, where the zero bias conductance linearly increases up to $T_c$ at which point $\Delta_{SC}$ vanishes, is reminiscent of the behavior of high-temperature superconductors with a nodal order parameter[35].

We extend our analysis to the entire superconducting dome. Figure 3f shows $dI/dV$ versus $V_{int}$ and $V_{TG}$. The doping dependence of the superconducting gap is already evident in the raw tunneling spectra. Taking advantage of the simultaneous transport measurements (Fig. 3g), we extract the superconducting gap $\bar{\Delta}_{SC}$ across the dome (see Methods). $\bar{\Delta}_{SC}$ closely follows the superconducting dome measured in transport, as shown by the pink diamonds in Fig. 3g (see Fig. S5 for the ratio $2\bar{\Delta}_{SC}/k_B T_{BKT}$ across the dome). This observation demonstrates that the presence of the superconducting gap $\bar{\Delta}_{SC}$ is directly related to the emergence of the superconducting phase coherence in MATTG, and once again confirms that the gap in the low-energy tunneling spectra is the superconducting gap. We note that in a recent study[20], the superfluid stiffness of MATTG was shown to exhibit a similar doping dependence to that of $T_c$. Together with our results, this indicates that both the superconducting gap and the superfluid stiffness scale with $T_c$, which can help further elucidate the nature of the superconductivity. All of the simultaneous measurements, analyses, and conclusions are well reproduced in the electron-doped superconducting phase, as shown in Fig. 4.

Interestingly, the higher-energy gap $\Delta_{HG}$ also exhibits a doping dependence across the superconducting domes, as shown in Figs. 3h and 4h, although it does not follow precisely the doping dependence of $T_{BKT}$. If one uses the Dynes formula with a nodal order parameter to fit this spectrum, the gap value is $\Delta_{HG} \approx 1$ meV near optimal doping of the superconducting phase, on the same order as the size of the tunneling gaps observed in previous STM measurements[13,14]. The ratio $2\Delta_{HG}/k_B T_{HG}$ is around 4. Moreover, if we use the critical temperature of the superconducting phase, $T_{BKT}$, the ratio $2\Delta_{HG}/k_B T_{BKT}$ is much higher, around 20 ~ 30 (see Fig. S5 for the ratios across the full dome). These observations suggest that $\Delta_{HG}$ may play an important role in the emergence of superconductivity, such as forming a relevant Fermi surface for the parent state of the superconducting phase, potentially related to the importance of flavor symmetry breaking[8,28–30].

**Volovik effect and gap-filling behavior with magnetic field**

A complementary way of probing the nature of the superconducting gap is by investigating the change in the DOS with a small out-of-plane or in-plane magnetic field, which had not been previously explored. Figure 5a shows $R_{xx,Top}$ as a function of $V_{TG}$ and $B_\perp$ for the hole-doped superconducting dome. The perpendicular critical magnetic field $B_{c,\perp}$ at which $R_{xx,Top}$ becomes nonzero is around 0.04 T at optimal doping. The corresponding superconducting tunneling spectra at optimal doping are shown in Fig. 5b. As previously observed in Fig. 2d, the depletion in the DOS near $V_{int} = 0$ V rapidly gets lifted once $B_\perp$ is applied. Figure 5c shows the zero bias conductance extracted from Fig. 5b. We find that the $B_\perp$-dependence of the zero bias conductance in our system exhibits a square-root dependence on $B_\perp$, as demonstrated by the fitting in Fig. 5c (see also Fig. S6 for the data on the electron-doped regime). This behavior aligns with the expected $\sqrt{B_\perp}$-dependence of the zero-energy DOS in a superconductor with a nodal order parameter, known as the Volovik effect[36,37].

Figure 5d shows $R_{xx,Top}$ versus $V_{TG}$ and in-plane magnetic field $B_{c,\parallel}$, and Fig. 5e shows the corresponding tunneling spectra at optimal doping up to $B_\parallel$ = 5 T. As soon as $B_\parallel$ is applied, the zero bias tunneling conductance starts increasing, which is shown together with the simultaneous transport measurements in Fig. 5f. This observation demonstrates that $dI/dV$ ($V_{int}$ = 0) starts rising at small magnetic field well below and up to $B_{c,\parallel} \approx 2$ T, above which the value saturates. Such gap filling behavior further supports a nodal order parameter for the superconducting phase. One possibility is a nodal spin-singlet superconductor, in which the zero bias DOS changes rapidly and increases linearly with the in-plane magnetic field magnitude[38,39], a behavior markedly different from that of conventional $s$-wave superconductors, where the DOS changes mostly when the Zeeman energy approaches the superconducting gap value. Another possibility is a nodal spin triplet or spin-valley-locked[40] order parameter with substantial in-plane orbital effects[41]. Given that the mirror symmetry of the t-MATTG layer is likely broken by the highly asymmetric environment, featuring a thick hBN dielectric on the top and atomically thin tunneling hBN with b-MATTG on the bottom, the in-plane orbital effects may not be negligible[10,41].

Our experimental setup can be generalized by modifying structural aspects, such as using a thinner tunnel barrier for the fabrication of vertical moiré Josephson junctions, which can further elucidate the superconducting order parameter of the system. Moreover, the versatility in performing simultaneous transport measurements and tunneling spectroscopy not only allows the investigation of superconductivity but also broadens the scope for studying the electronic structure of moiré materials, including air-sensitive materials like twisted transition metal dichalcogenides, which can be challenging to study via scanning-probe techniques that require open surfaces.

**Acknowledgements:** We thank Ady Stern, Ali Yazdani, Allan MacDonald, Ashvin Vishwanath, Eslam Khalaf, Kevin Nuckolls, Liang Fu, Paco Guinea, Patrick Ledwith, Pavel Volkov, Senthil Todadri, and Steve Kivelson for fruitful discussions.

This work has been primarily supported by the Army Research Office MURI W911NF2120147; with support also by the 2DMAGIC MURI FA9550-19-1-0390, the MIT/Microsystems Technology Laboratories Samsung Semiconductor Research Fund, the Sagol WIS-MIT Bridge Program, the National Science Foundation (DMR-1809802), the Gordon and Betty Moore Foundation's EPiQS Initiative through grant GBMF9463, and the Ramon Areces Foundation (to P.J.H.). K.W. and T.T. acknowledge support from the JSPS KAKENHI (Grant Numbers 21H05233 and 23H02052) and World Premier International Research Center Initiative (WPI), MEXT, Japan. This work made use of Harvard's Center for Nanoscale Systems, supported by the NSF (ECS-0335765).

**Author contributions:** J.M.P. designed the experiment. J.M.P. and S.S. fabricated devices and conducted measurements. K.W. and T.T. provided hBN samples. J.M.P., S.S., and P.J.-H. performed data analysis and wrote the manuscript with contributions from all authors.

**Competing interests:** The authors declare no competing interests.

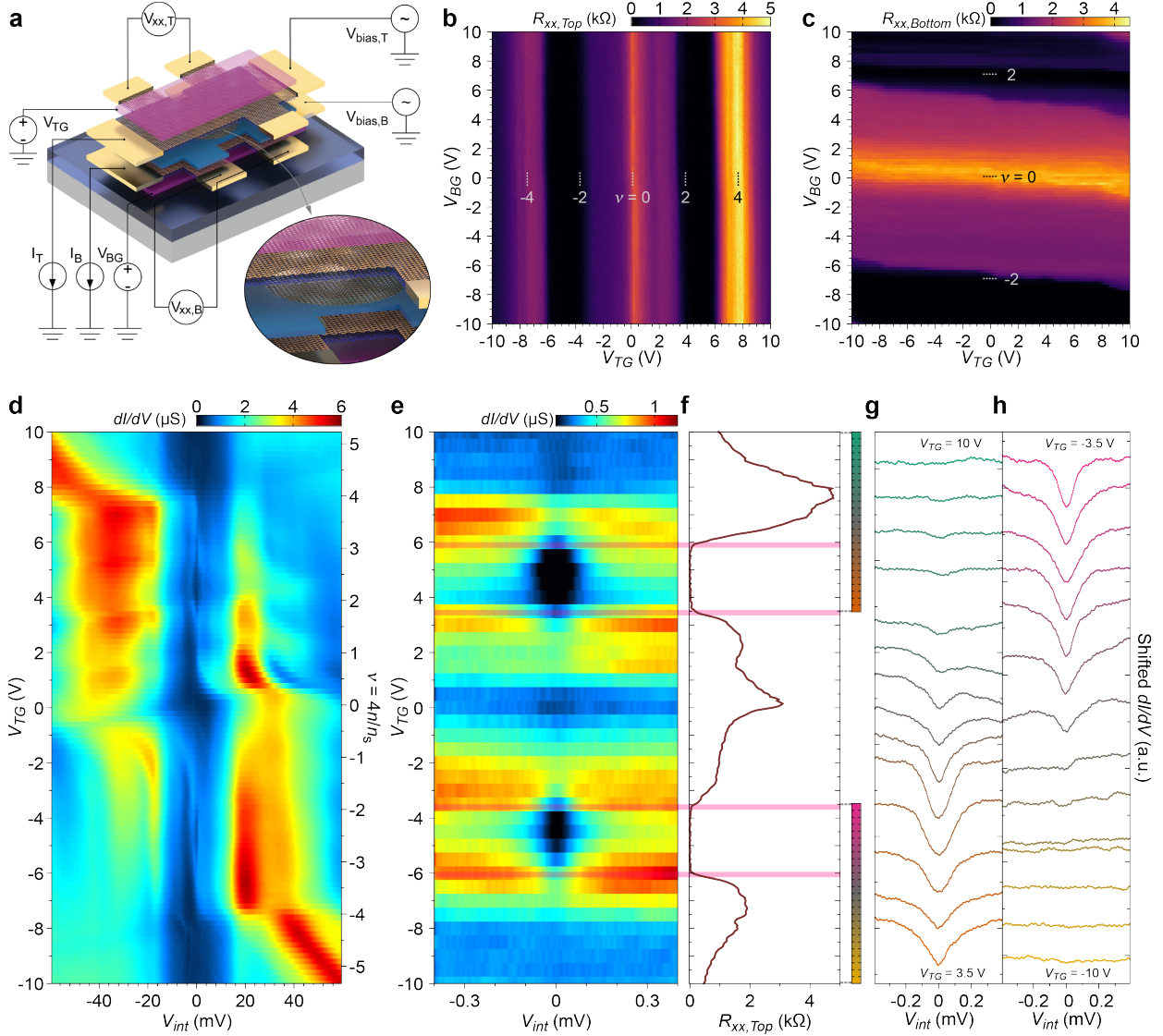

**Fig. 1. Simultaneous tunneling spectroscopy and transport measurements.** (a) An illustration of the device structure. Two magic-angle twisted trilayer graphene (MATTG) layers are coupled through a thin hBN tunnel barrier. Controllable tunneling spectroscopy is enabled by an additional hBN spacing layer with a hole, restricting electron tunneling to the core region of the device to align with the transport measurements (see inset for zoom-in). Each MATTG is shaped into a Hall bar with independent electrostatic gating, allowing simultaneous transport measurements. (b-c) Resistance as a function of top and bottom gate voltages ($V_{TG}$ and $V_{BG}$) for top-MATTG (t-MATTG), $R_{xx,Top}$ (b), and bottom-MATTG (b-MATTG), $R_{xx,Bottom}$ (c). Filling factor $v = 4n/n_s$, where $n$ is the carrier density and $n_s = 8\theta^2/(\sqrt{3}a^2)$ is the superlattice density ($a = 0.246$ nm is the graphene lattice constant), is annotated. (b) For t-MATTG, $v = \pm 4$ appears around $V_{TG} = \pm 7.5$ V (for $V_{BG} = 0$ V). Superconductivity is observed at $V_{TG} = -6.1$ V $\sim -3.5$ V and $3.4$ V $\sim 6$ V on the hole-doped and electron-doped sides, respectively. The twist angle for t-MATTG is $\theta_T \approx 1.42°$. (c) b-MATTG exhibits superconducting phases starting from $\sim \pm 6.5$ V (for $V_{TG} = 0$ V). The twist angle for b-MATTG is $\theta_B \approx 1.55°$. (d) Tunneling conductance $dI/dV$ as a function of interlayer bias voltage $V_{int}$ and $V_{TG}$ at $T \approx 200$ mK. b-MATTG is tuned to a normal metallic phase ($V_{BG} = -3$ V). The typical cascade of phase transitions is observed (see main text). Finer features, including gaps

centered around zero bias, develop at small $V_{int}$. (e) A zoom-in of $dI/dV$ near zero bias. (f) $R_{xx,Top}$ measured for direct comparison. The doping ranges for superconductivity with zero $R_{xx,Top}$ coincide with the ranges where $dI/dV$ develops pronounced dips around zero bias (pink lines), suggesting that these features originate from the superconducting phases. (g-h) Line cuts of the tunneling spectra in the doping ranges specified by the color bars matching the y-axis of d-f. V-shaped gaps develop only in the presence of superconductivity.

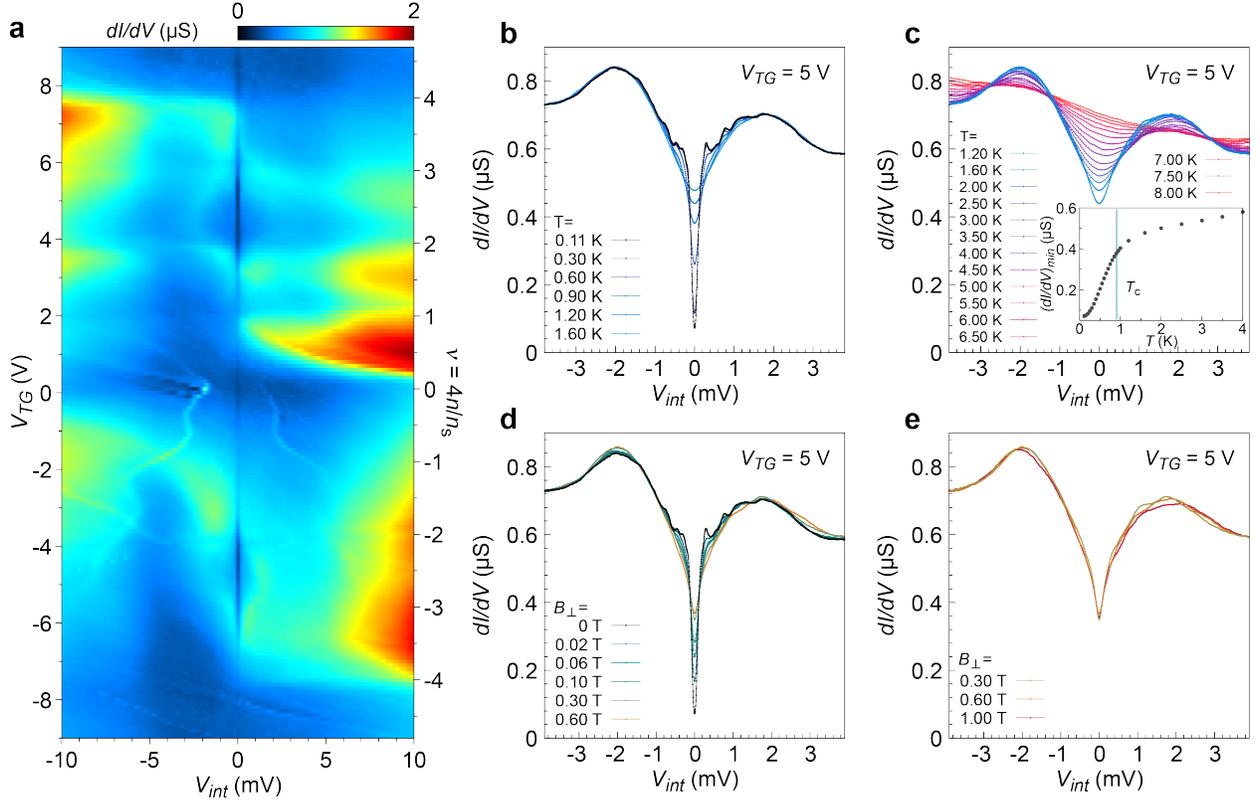

**Fig. 2. Coexistence of two distinct energy gaps**. (a) $dI/dV$ versus $V_{int}$ and $V_{TG}$ at $T = 110$ mK ($V_{BG} = -3$ V). In addition to the V-shaped gaps identified in Figs. 1g and 1h, a depletion of DOS, albeit with an energy scale of a few meV, is visible in the doping ranges where superconductivity appears (Fig. 1f). (b) Temperature dependence of $dI/dV$ in the optimally-doped superconducting phase on the electron-doped side ($V_{TG} = 5$ V). At $T = 110$ mK, a low-energy gap appears as a sharp, deep feature on a scale of 0.1 meV, whereas a higher-energy gap ($\Delta_{HG}$) manifests as a broader, shallower depletion of DOS on a scale of 1 meV. As the temperature increases, the low-energy gap rapidly vanishes around the superconducting critical temperature $T_c$ measured in transport ($\sim 1$ K), while $\Delta_{HG}$ persists. (c) At higher temperatures, the $\Delta_{HG}$ spectrum fills up slowly and becomes featureless above $T = 8$ K. Inset: minimum tunneling conductance $(dI/dV)_{min}$ versus $T$ shows a rapid increase up to $\sim T_c$ and a transition to a slower increasing rate above $T_c$ (dashed line). (d-e) Perpendicular magnetic field ($B_\perp$) dependence of $dI/dV$. The low-energy gap fills up rapidly with small $B_\perp$ on the scale of the critical magnetic field ($B_{c,\perp}$) measured in transport (see Fig. 5 and Fig. S6), whereas $\Delta_{HG}$, in contrast, remains largely unaffected even up to $B_\perp = 1$ T, much higher than $B_{c,\perp}$. This trend is consistent on the hole-doped side as well. Therefore, we conclude that the low-energy gap is the superconducting gap, and the higher-energy gap is a separate gap, whose coexistence is observed for the first time in tunneling spectroscopy in a moiré superconductor. For b-e, b-MATTG is set to a normal metallic phase ($V_{BG} = 4$ V). We note that there is no qualitative difference in using different $V_{BG}$ in the metallic range, and our choice of $V_{BG}$ exhibits a flat DOS and no temperature dependence (see Figs. S1 and S3).

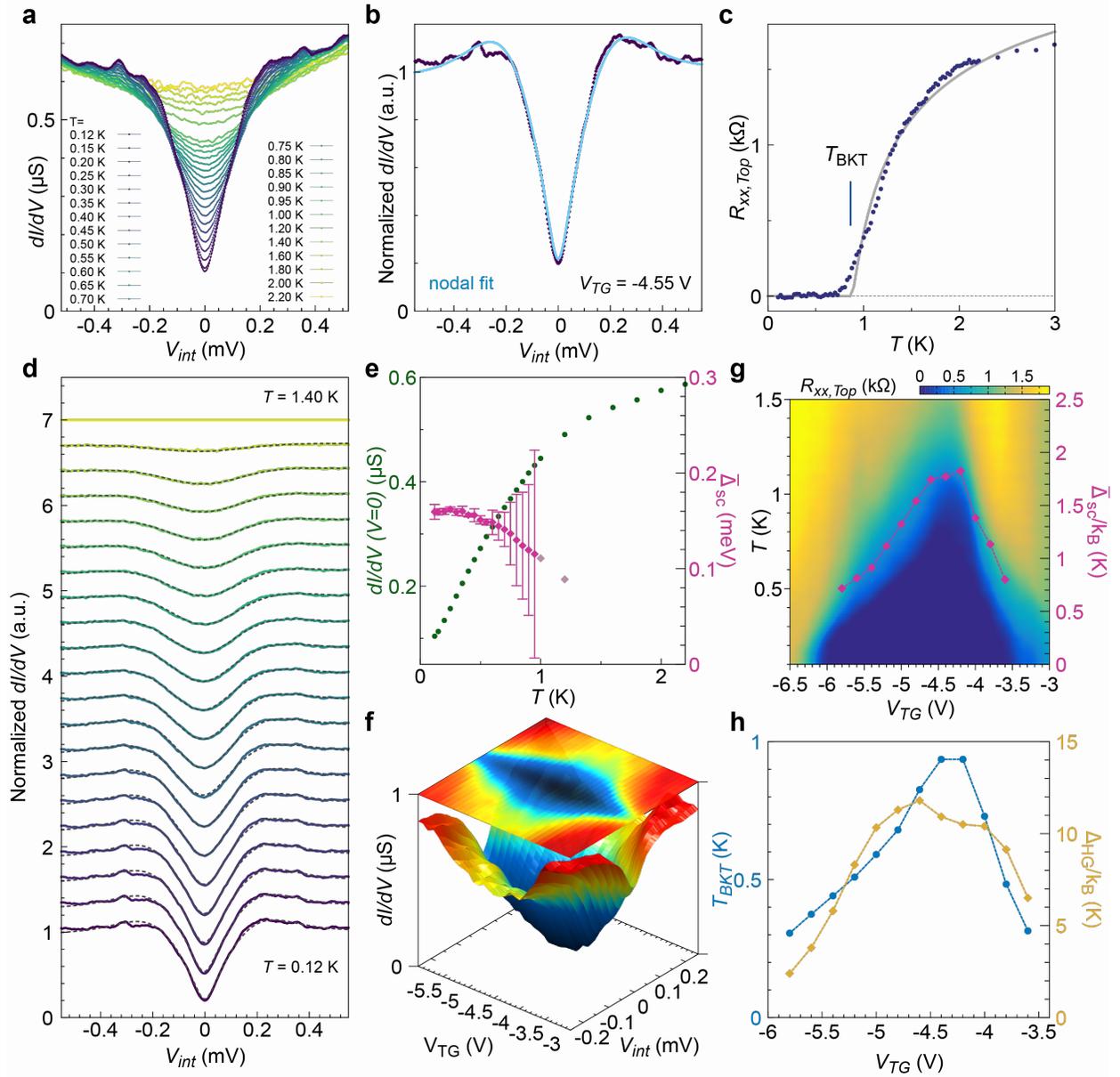

**Fig. 3. Gap filling behavior of tunneling spectra and doping dependence of the superconducting gap for hole-doped superconductivity.** (a) Raw $dI/dV$ versus $V_{int}$ at $V_{TG} = -4.55$ V, near optimal doping, as a function of temperature up to $T = 2.2$ K. (b) Normalized $dI/dV$ spectrum at $T = 120$ mK, with a normalization temperature of 1.4 K, taken slightly above $T_c$ (see Methods for definition) based on the simultaneous transport measurements. The light blue line shows the fit to the modified Dynes formula with a nodal order parameter and a gap distribution (see Methods for fitting parameters and Fig. S4 for comparison with fits using an *s*-wave order parameter). The extracted mean superconducting gap value is $\bar{\Delta}_{SC} = 0.159$ meV. (c) $R_{xx,Top}$ versus $T$ reveals the Berezinskii–Kosterlitz–Thouless (BKT) transition temperature, $T_{BKT} \approx 0.86$ K. The normal to superconductor transition starts around $T = 2$ K. (d) Normalized $dI/dV$ spectra versus $T$ up to 1.4 K. Dashed lines show the modified Dynes formula fitting results. (e) Zero bias conductance $dI/dV$ ($V_{int} = 0$) versus $T$ (green, left axis) reveals a linear filling behavior starting from the lowest temperature, far below $T_c$. The figure also shows $\bar{\Delta}_{SC}$ as a function of $T$ (pink, right

axis), which decreases slowly as $T$ increases, with a vanishing trend around $T_c$. The error bars denote 95% confidence intervals. The grey values have large confidence intervals due to the increased uncertainty in gap fitting near $T_c$. (f) $dI/dV$ versus $V_{int}$ and $V_{TG}$ reveals the variation of the tunneling spectra across the superconducting dome. (g) Doping dependence of $\bar{\Delta}_{SC}$ (pink, right axis) overlaid on $R_{xx,Top}$ versus $T$ and $V_{TG}$. The variation of $\bar{\Delta}_{SC}$ closely follows the shape of the superconducting dome as measured in transport. (h) $T_{BKT}$ (light blue, left axis) and $\Delta_{HG}$ (yellow, right axis) versus $V_{TG}$, showing that they have a similar doping dependence.

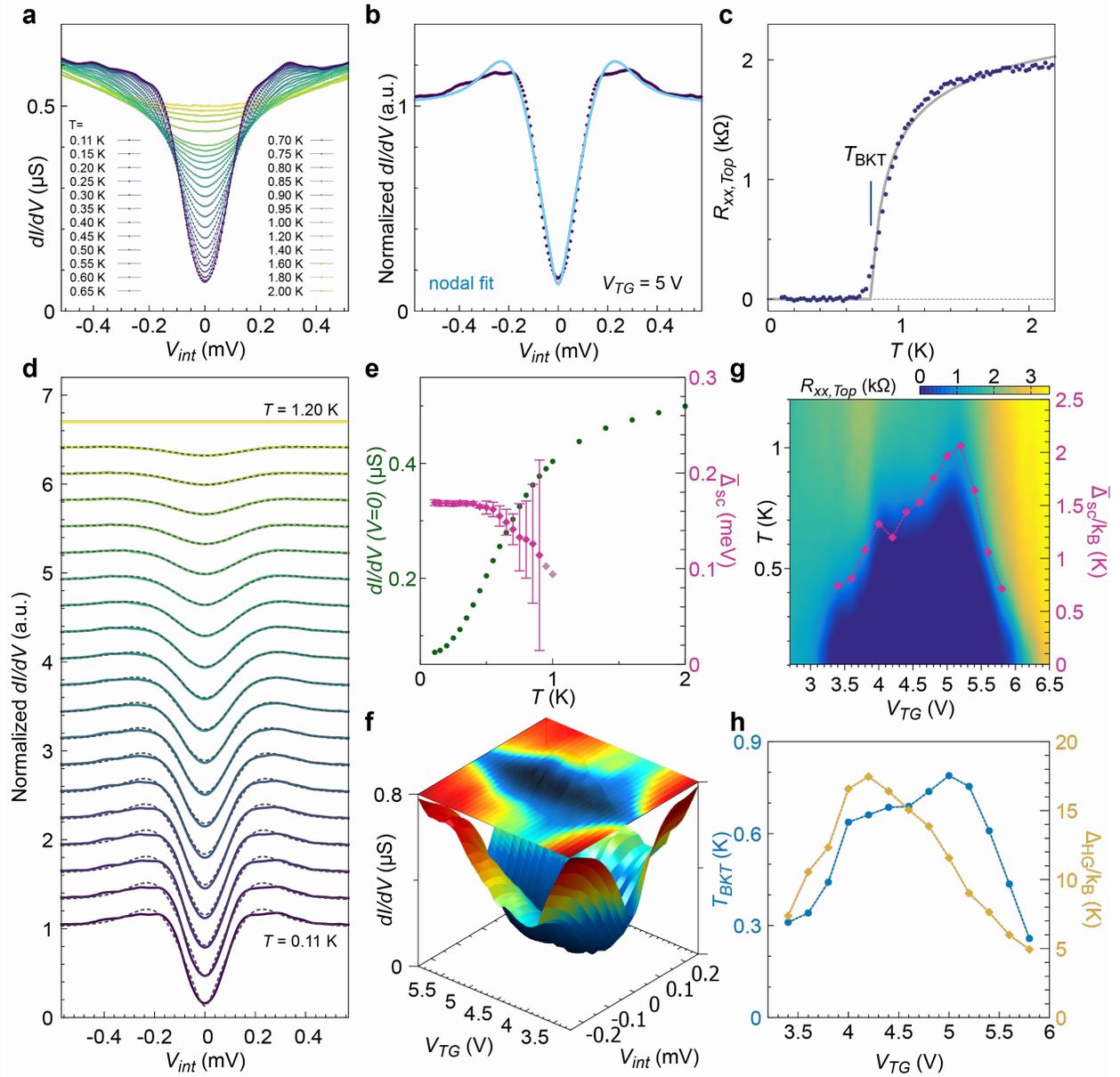

**Fig. 4. Gap filling behavior of tunneling spectra and doping dependence of the superconducting gap for electron-doped superconductivity.** (a) Temperature dependence of $dI/dV$ versus $V_{int}$ near optimal doping ($V_{TG}$ = 5 V), as a function of temperature up to $T$ = 2 K. (b) Normalized $dI/dV$ spectrum at $T$ = 110 mK (normalized by the 1.2 K spectrum, slightly above $T_c$). The light blue line shows the fit to the modified Dynes formula, with $\bar{\Delta}_{SC}$ = 0.17 meV. (c) Temperature dependence of $R_{xx,Top}$ reveals $T_{BKT} \approx$ 0.79 K, and the normal to superconductor transition starts around $T$ = 1.5 K. (d) Temperature dependence of the normalized $dI/dV$ spectra up to $T$ = 1.2 K. Dashed lines show the modified Dynes formula fitting results. (e) Temperature dependence of the zero bias conductance $dI/dV$ ($V_{int}$ = 0) (green, left axis) reveals the low-temperature linear filling behavior. The figure also shows nodal gap amplitude $\bar{\Delta}_{SC}$ as a function of $T$ (pink, right axis). Error bars denote 95% confidence intervals. Note that the grey values have large confidence intervals as they are close to $T_c$. (f) Doping dependence of $dI/dV$ illustrates the evolution of the tunneling spectra across the superconducting dome. (g) Doping dependence of

$\bar{\Delta}_{SC}$ (pink, right axis) is overlaid on $R_{xx,Top}$ versus $T$ and $V_{TG}$. The trend of $\bar{\Delta}_{SC}$ closely follows the shape of the superconducting dome as measured in transport. (h) $T_{BKT}$ (light blue, left axis) and $\Delta_{HG}$ (yellow, right axis) versus $V_{TG}$, showing that they have a similar doping dependence (though the maxima occur at different dopings).

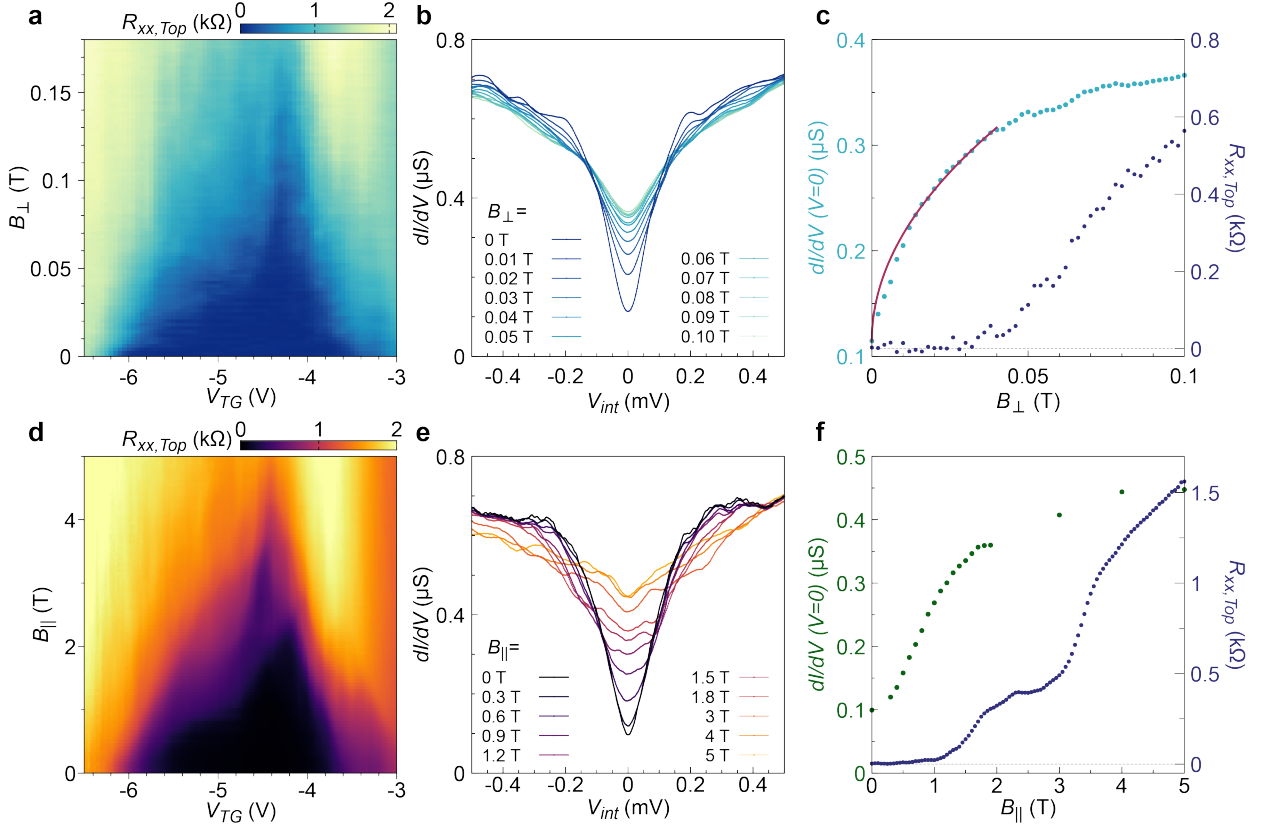

**Fig. 5. Out-of-plane and in-plane magnetic field dependence of the superconducting phase.**
(a) $R_{xx,Top}$ versus $B_\perp$ and $V_{TG}$ shows the superconducting dome on the hole-doped side. (b) $B_\perp$-dependence of $dI/dV$ at hole-doped optimal doping ($V_{TG}$ = −4.55 V). Application of small $B_\perp$ leads to rapid filling of the superconducting gap. (c) Zero bias conductance $dI/dV$ ($V_{int}$ = 0) versus $B_\perp$ (light blue, left axis), extracted from (b), and $R_{xx,Top}$ versus $B_\perp$ (dark blue, right axis). The red line is obtained by fitting $dI/dV$ ($V_{int}$ = 0) below the critical magnetic field $B_{c,\perp}$ ~ 0.04 T with $A \cdot \sqrt{B_\perp} + G_0$, where A is a phenomenological factor and $G_0$ denotes the $dI/dV$ ($V_{int}$ = 0) value at zero magnetic field, consistent with the predicted Volovik effect for a nodal superconductor. (d) In-plane magnetic field $B_\parallel$ and doping dependence of $R_{xx,Top}$. (e) $dI/dV$ versus $V_{int}$ up to $B_\parallel$ = 5 T around optimal doping ($V_{TG}$ = −4.55 V). (f) $dI/dV$ ($V_{int}$ = 0) versus $B_\parallel$ (green, left axis), extracted from (e), and $R_{xx,Top}$ versus $B_\parallel$ (dark blue, right axis). The zero bias conductance increases immediately upon application of small $B_\parallel$. The slope becomes smaller around the in-plane critical magnetic field $B_{c,\parallel}$ ≈ 2 T. We note that the double step seen in d and f could have different origins, such as inhomogeneity or a pseudogap state. For all data in this figure, b-MATTG is set to a metallic phase ($V_{BG}$ = −3 V).